\documentclass[twocolumn]{aastex62}

\usepackage{graphicx}
\usepackage{epstopdf}
\usepackage{natbib}
\usepackage{amssymb}
\usepackage{amsmath}
\usepackage{float}
\usepackage{chngcntr}

\newcommand{\beq}{\begin{equation}}
\newcommand{\eeq}{\end{equation}}
\newcommand{\simm}{${\sim}$}

\def\mj{M$_\mathrm{J}$}
\def\rj{R$_\mathrm{J}$}

\received{2019 September 20}
\revised{2019 December 7}
\accepted{2020 January 2}
\submitjournal{The Astronomical Journal}

\shorttitle{M4 Planet Occurrence} 
\shortauthors{Wallace, Hartman, Bakos et al.}

\begin{document}

\title{A Search for Transiting Planets in the Globular Cluster M4 with K2: Candidates and Occurrence Limits}

\author[0000-0001-6135-3086]{Joshua J. Wallace}
\affiliation{Department of Astrophysical Sciences, Princeton
  University, 4 Ivy Ln, Princeton, NJ 08544, USA}

\author[0000-0001-8732-6166]{Joel D. Hartman}
\affiliation{Department of Astrophysical Sciences, Princeton
  University, 4 Ivy Ln, Princeton, NJ 08544, USA}

\author[0000-0001-7204-6727]{G\'asp\'ar \'A. Bakos}
\altaffiliation{MTA Distinguished
Guest Fellow, Konkoly Observatory}
\affiliation{Department of Astrophysical Sciences, Princeton
  University, 4 Ivy Ln, Princeton, NJ 08544, USA}

\correspondingauthor{Joshua Wallace}
\email{joshuawallace800@gmail.com}

\begin{abstract}
We perform a search for transiting planets in the NASA K2 observations of the globular cluster (GC) M4.  This search is sensitive to larger orbital periods ($P\lesssim 35$ days, compared to the previous best of $P\lesssim 16$ days) and, at the shortest periods, smaller planet radii (R$_p\gtrsim0.3$ \rj, compared to the previous best of R$_p\gtrsim0.8$ \rj) than any previous search for GC planets.  Seven planet candidates are presented.  An analysis of the systematic noise in our data shows that most, if not all, of these candidates are likely false alarms.  We calculate  planet occurrence rates assuming our highest significance candidate is a planet and occurrence rate upper limits assuming no detections. We calculate 3$\sigma$ occurrence rate upper limits of 6.1\% for 0.71--2 \rj\ planets with 1--36 day periods and 16\% for 0.36--0.71 \rj\ planets with 1--10 day periods. The occurrence rates from {\it Kepler}, {\it TESS}, and RV studies of field stars are consistent with both a non-detection of a planet and detection of a single hot Jupiter in our data. Comparing to previous studies of GCs, we are unable to place a more stringent constraint than \cite{gilliland2000} for the radius--period range they were sensitive to, but do place tighter constraints than both \cite{weldrake2008} and \cite{nascimbeni2012} for the large-radius regimes to which they were sensitive.  
 \end{abstract}

\keywords{Exoplanets, Globular star clusters, Hot Jupiters, Transit photometry}

\section{Introduction}
\label{sec3:intro}

The globular cluster (GC) M4 (NGC 6121) was observed by the K2 mission
(\citealt{howell2014}) 
during its Campaign 2.  These data underwent a blanket search for
variable objects in our previous work (\citealt{wallacem4}), but did not receive
a focused search for planetary transits.  Any constraints that could be put on
planet occurrence rates in a GC would be of scientific interest. 
GCs provide more-or-less homogeneous populations of
metal-poor stars---M4 in particular has a metallicity of [Fe/H]${\approx}-1.2$
(\citealt{harriscatalog}, 2010 edition).  As such, they would provide
valuable test beds for 
theories about planet formation and its dependence on stellar
metallicities 
(e.g., \citealt{ida2004,johansen2009,ercolano2010,johnson2010,johnson2012}),
assuming such formation mechanisms also take into account the
denser stellar environment.
These high stellar densities could also provide a fruitful
testbed of dynamical planet formation and evolution 
theories.  The relatively 
large number of stellar encounters in GCs, due to both their old ages and high
stellar densities, are thought to kick planets out of planetary
systems, especially those on wide
orbits \citep{sigurdsson1992,davies2001,bonnell2001,fregeau2006,spurzem2009}.
However, stellar encounters are 
also expected to increase the probability of 
formation of hot Jupiters (HJs) via high-eccentricity migration in some cases (\citealt{hamers2017}).
For reference for this work, a typical definition of an HJ is a planet
with a radius ${\gtrsim}0.8$ \rj\ and an orbital period ${\lesssim}10$
days. 
An enhanced HJ
occurrence rate in GCs may suggest a preference for dynamical
formation pathways in our own neighborhood for close-in planets.  And
finally, HJs are expected to undergo tidal orbital decay on Gyr
timescales (e.g., \citealt{penev2018}), and the old ages of GCs may be
helpful in testing this theoretical expectation.

M4 holds the distinction of possessing the only planet known in a GC,
PSR 1620-26~b, a  planetary-mass object orbiting a pulsar--white dwarf binary (\citealt{backer1993,thorsett1993,michel1994,rasio1994,arzoumanian1996,thorsett1999,ford2000,richer2003}), with its mass measured by \cite{sigurdsson} to be $2.5\pm1$ M$_J$.
Work since its discovery has shown that this
planet may have formed later in the cluster's life, rather than
from a protostellar disk. For example, \cite{beer2004} propose a model where a
stellar encounter during the common envelope phase that led to the
formation of the pulsar caused a dynamical instability in the dense
equatorial wind formed as part of the common envelope phase. As such,
PSR 1620-26~b may not 
be able to provide constraints on planet formation processes in GCs
that are contemporaneous with star formation.  

Given the relatively large distance to M4 of
1.8 kpc (\citealt{hendricks2012,kaluzny2013b,braga2015,neeley2015}),
a wide-scale radial velocity (RV) survey to search for planets is
impractical, but a photometric survey to search for transits is
feasible.  Despite its distance, M4 is nevertheless the closest GC, and it has a
relatively sparse core, so it is
perhaps the best target for discovering additional GC planets.

Previous searches for transiting planets have been made  in GCs.
The largest to
date are the {\it 
HST} campaign of \cite{gilliland2000} and the ground-based campaign of
\cite{weldrake2005}, both searching for planetary transits in
47 Tuc.  \cite{gilliland2000} state that the reason for choosing this cluster (its distance is about twice that of M4) is because its spatial and main-sequence brightness distributions matched well the capabilities of {\it HST}. They obtained high-precision photometry for ${\sim}$34,000
stars over an 8.3 day observing campaign.  With the then-current
understanding of HJ occurrence rates, they had expected to find
${\sim}$17 planets; however, they found none.  More
recently, \cite{masuda2017} revised the expected number of planets
that Gilliland et al.~would have found to $2.2^{+1.6}_{-1.1}$, based on an
updated understanding of planet occurrence from the {\it Kepler} mission.
This revised number additionally does not account for the lower
metallicity of 47 Tuc ([Fe/H]${\approx}-0.7$; \citealt{mcwilliam2008})
relative to the {\it Kepler} stars, which are primarily field stars.
This is expected to revise the number even 
lower due to the metallicity dependence of the occurrence rate of
HJs  (\citealt{santos2001,fischer2005,petigura2018}). 
\cite{weldrake2005} used the Australian National University 40 in telescope
at Siding Spring observatory over 33 nights and obtained a much wider
field of view than Gilliland et al.'s {\it HST} observations,
observing out to 60\% of 47 Tuc's tidal radius.  They
obtained light curves for ${\sim}$110,000 stars (though only
${\sim}$20,000 of these had sufficiently low scatter to be sensitive
to HJs) and could detect giant planets with 
periods up to 16 days.  Their calculated expected planet yield was
${\sim}$7 planets (based on 1 \rj\ planets with periods less than 16
days and an intrinsic formation rate of 0.8\%), but they found none.
\cite{masuda2017} revised the expected number of planets from this survey down by about a factor of four.
Both the \cite{gilliland2000} and \cite{weldrake2005} surveys were not sensitive to much other than HJs.
Other searches for transiting exoplanets in GCs include the search
of \cite{weldrake2008} in $\omega$ Cen, which for the most part was
sensitive only to planets with radii ${>}1.5$ \rj\ and had no detections,
and \cite{nascimbeni2012} in NGC 6397, which with a null detection and 
${\sim}5000$ light curves, were not able to derive constraints on planet
occurrence that fell below the occurrence rates measured by {\it Kepler}.

M4 was in the field of
view of the {\it Kepler} telescope during Campaign 2 (running from 2014 August 23 to 2014 November 10) of the K2 mission
 and, as mentioned, 
continuous observations of a
portion of this cluster were included in the data downloaded from the
telescope.  Though the original proposals to obtain these data were
focused on observing RR Lyrae variables in the cluster, the excellent
photometry and long-term coverage of M4 allows for  detecting
other variable objects, potentially  including 
transiting HJs. The {\it Kepler} telescope and detector were not
designed with GC observations in mind: the \simm4\arcsec/pixel image resolution
 leads to significant blending in the images, and
the periodic telescope drift experienced during the K2 mission
produces systematic noise in the photometry.  

Despite these
problems, the longer observation span of K2 Campaign 2 relative to the
previous GC transiting planet searches potentially opens a new regime
of planetary orbital period to explore and  to place constraints on GC planet occurrence.  The
longer observations also increase the number of observed transits for
orbits of a given period relative to the previous surveys, helping to boost sensitivity to smaller
radius planets in the period ranges that have already been probed by
other GC studies.   
Given the scientific motivation for finding planets
in GCs and the new parameter space opened for exploration by these data,
and since the reduced data have scientific utility in addition to permitting a transit search  (see our variable catalog in
\citealt{wallacem4}), there is more than sufficient merit to motivate the effort for the
search. We summarize our photometric reductions and explain our
transit search methodology in Section~\ref{sec3:method}, present the
results of our transit search in Section~\ref{sec3:transitsearch}, and
provide planet occurrence rates and limits in
Section~\ref{sec3:occurrence}.  We then discuss the results in
Section~\ref{sec3:discussion} and conclude in Section~\ref{sec3:conclusion}.

\section{Method}
\label{sec3:method}

\subsection{Photometric Processing}
Our photometric processing pipeline is fully described in \cite{wallacem4},
and is similar to  that of \cite{soaresfurtado}.  We provide a brief
summary here.  M4 was observed by K2 for \simm 79 days
in 2014, during the mission's Campaign 2.  Given the high degree of blending in the images, we decided to use image subtraction \citep{imagesubtraction}
to extract light curves for the objects.  The {\it Gaia} first data
release (DR1) source
catalog (\citealt{gaiadr1,gaiamission}) was used as an astrometric and
photometric reference catalog.  We included all 
sources with 
$G<19$. DR1 was used instead of the {\it
  Gaia} second data release (DR2) catalog owing to our beginning this
study prior to {\it Gaia} DR2. Using the DR1 catalog leads to
  slightly higher upper limits on planet occurrence than could have
  been derived with the additional sources in DR2, however the
  improvement that could be 
  expected from using DR2 is too small to justify the additional
  effort of redoing the photometric reduction.
 In using {\it Gaia} DR1 as a photometric
reference catalog, we had to convert from $G$ to our {\it Kepler}
instrumental magnitudes.  We found that a simple additive conversion
was all that was needed, likely due to the similar bandpasses of the
two instruments, with the conversion from instrumental
  magnitude $M_\mathrm{I}$ being $G = M_\mathrm{I}+ 25.275$
  mag.  Any error in the reference magnitudes would lead to errors in
  the measured transit depths, but not in the significance of any
  signals as all of the light curve points would be affected equally.

The photometry was extracted for apertures of 1.5,
1.75, 2.0, 2.25, 2.5, 2.75, and 3.0 pixels in radius. At the end of our processing, we
found which aperture radius minimized the photometric scatter as a
function of $G$ and used the data from the corresponding aperture for
all objects of a given magnitude throughout our analysis.  For
reference, the typical full-width at half maximum value for the
point spread function of the images was \simm 1.5 pixels.

After extracting this raw photometry, systematic variability from the
spacecraft roll was cleaned up based our implementation of  the
algorithm developed by \cite{vanderburg2014} and \cite{vanderburg2016}.  The
light curves were then further cleaned of common systematic trends
using the trend filtering algorithm (TFA; \citealt{kovacs-tfa}) as
implemented in \texttt{VARTOOLS} (\citealt{vartools}).
For each aperture, 250 light curves were selected from uniform
  bins of image position and magnitude to be the trend light curves for
  the TFA. The total number of light curves thus produced was 4554.
In this work, all objects are referred to by the identifiers assigned them in \cite{wallacem4}; see table 1 in that work.
All of the raw and processed light curves
 are available at
 \cite{lightcurves}\footnote{Published at Princeton University's
   DataSpace and licensed under a Creative Commons Attribution 4.0
   International License, accessible at
   \url{http://arks.princeton.edu/ark:/88435/dsp01h415pd368}}. 

\subsection{Transit Search}

All 4554 light curves were searched for planet
transits 
using the \texttt{VARTOOLS} implementation of the 
box-fitting least squares \citep[BLS;][]{kovacs2002} algorithm. The
light curves were sigma clipped prior to the search (5$\sigma$, three
iterations). We ran some injected transits through our pipeline to
ensure that they would be recoverable even with the photometric post-processing
and sigma clipping.  
We searched periods between one day and the
maximum 
observation length of the given light curves (most having the maximum
length of \simm 78 days, which is slightly shorter than the full span of observations owing to our need to trim the first \simm1 day of data).  
A one day orbit around the most massive and evolved cluster-member
stars under consideration (0.81 M$_\sun$) has an orbital semimajor
axis of 3.9 R$_\sun$, compared to a stellar radius of 4.9 R$_\sun$ for
these same stars.  Thus for the handful of the most evolved stars
under consideration, one day orbits are not possible, but they are possible for the vast
majority of the stars we consider.
 
The range of values for the fractional
transit duration $q$ used in the search varied between $0.1\times
q_\mathrm{exp}$ 
and $2 \times q_\mathrm{exp}$, with $q_\mathrm{exp}$ being the
expected transit duration at a given period based on the density of the
given star (see the next two paragraphs), assuming a circular orbit, and that the impact parameter $b$ is zero.
The minimum $q$ searched was adjusted as necessary so that it
was never less than $(\Delta t)_\mathrm{min}/P$, with $(\Delta
t)_\mathrm{min}$ being the minimum time between observations (the
$Kepler$ cadence, 29.4 minutes, adjusted slightly based on the actual BJD values of the observations) and $P$ being the period being
searched. 
The number of phase
bins used was 
set to $2/q_\mathrm{min}$, with $q_\mathrm{min}$ here being
$0.1\times q_\mathrm{exp}$, up to a maximum value of 2100.  This value is based on
the expected transit duration of the shortest-period planets around
the smallest-radius stars in our sample.  With having less than 4000
measurements per light curve, a greater number of phase bins  would
not have been useful anyway.

Stellar densities were calculated based on an isochrone fit
for the cluster and the $V$ magnitudes taken from
\cite{mochejska2002}.
This
isochrone-based density determination  
produced incorrect densities  for objects that were not
cluster members, but since they were not included in our final
analysis this did not matter.
For objects not found in the \cite{mochejska2002} catalog, their $V$ magnitudes were 
converted from $G$ based on a second-degree polynomial fit between $G$
and the \cite{mochejska2002} $V$ magnitudes of the matched objects. 
The
isochrone was taken from the calculations of \cite{yi2001}.
  For the isochrone, we assumed  a 
metallicity of [Fe/H]$=$ -1.2 (\citealt{harriscatalog}, 2010 edition)
and an estimated age of 11.3 Gyr based on our fit to the data.
The $V$ magnitudes were used instead of {\it Gaia} $G$ because this
calculation was performed earlier in {\it Gaia}'s mission and our
isochrone database had not yet incorporated results that used the {\it
  Gaia} bandpass.
The $V$
magnitudes were converted to absolute magnitudes assuming a cluster
distance of 1.8 kpc
(\citealp{hendricks2012,kaluzny2013b,braga2015,neeley2015}) and 
an $A_V$ extinction of 1.24 mag based on our fit to the data, a value
\simm 0.15 less than the mean value found by \cite{hendricks2012}.
We also note that there is differential reddening across the cluster,
with  \cite{hendricks2012} finding the difference between the lowest
and the highest $E(B-V)$ values to be ${\sim}0.2$ mag.

The use of stellar density in the BLS calculation tailors the transit
duration range that is searched to those expected for a planet around
a star of the given density.  This boosts sensitivity to physically
likely transit scenarios. This is in contrast to our previous search
in \cite{wallacem4}, which was not restricted to only planetary transit-minded transit durations, and is part of the reason we were better able to
find planet candidates in this search (the previous search
found none). Given the fairly wide distribution in
$q$ that we search ($0.1\times q_\mathrm{exp}$ to $2 \times
q_\mathrm{exp}$), the search is insensitive to modest errors in
density. 

The phase-folded light curves and periodograms for the periods of the top five periodogram peaks were then examined by eye for
significance.  
 We used the
\texttt{checkplot} module of \texttt{astrobase} (\citealt{astrobase})
for this by-eye examination.
As part of this, obvious non-planet-transit physical signals were
excluded (e.g., RR Lyrae and eclipsing binaries, as well as objects
blended with them).  We also found that a large and temporary
 systematic variation in brightness  
occurred about halfway through the observation in the light curves of
many of the objects, at the point in the campaign when the
roll direction of the spacecraft changed.  This light curve systematic in some cases 
phased up with 
other outliers to produce large-period BLS signals; cases where this
happened 
were determined by eye and removed.  In ambiguous cases, we erred on
the side of completeness and 
included the objects in our subsequent consideration, since we did not
want to impose an unquantified limit on signal-to-pink-noise (S/PN) that was stricter than the
hard limit that we used---only those objects with a
S/PN greater than eight were examined by eye.
Approximately half of the total number of objects passed this limit. We wish to note that, at this
stage, all of the objects were considered without respect to their cluster
membership status. Although our initial S/PN
threshold was set to eight, we found later that a higher threshold
should be used, as detailed subsequently in Section~\ref{sec:spnthreshold}.

\subsection{Signal-to-pink Noise Threshold Determination}
\label{sec:spnthreshold}

Given the residual systematic noise left in our data---largely leftover
from the roll-correlated variability that was mostly but not entirely
removed by our processing pipeline---we found S/PN to be a useful
metric in evaluating signal significance.  We used the S/PN as
calculated by \texttt{VARTOOLS}, which is based on the definition of
\cite{pont2006}. The signal value used in this calculation is the BLS
transit depth and the pink noise is a quadrature sum of the light
curve white noise divided by the number of points in transit and the
light curve red noise (calculated from the RMS of the binned light
curve with bin size equal to the transit duration) divided by the
number of transits.  After an initial search through the BLS search
results, we saw transit-like signals that had lower S/PN than
would be expected based on the observed white noise, suggesting a
significant amount of correlated noise that could mimic transits.  To
better understand how well the correlated noise could mimic transits in
our data, we reran our BLS search with the same parameters as before,
but this time looking instead for ``anti-transits'', 
periodic box-shaped brightenings in the data instead of dimmings.
Since there are no common periodic astrophysical phenomena that can produce
such brightening signals at the \simm 10 mmag level that Jupiter-sized planets
produce for dimmings, these presumably are all due to
noise. Gravitational self-lenses from binary systems consisting of a
neutron star/black hole and a main sequence star can produce such
signals, but the occurrence rates for such objects are expected to be
low (e.g., \citealt{masuda2018} expect the {\it TESS} survey to
 have a detection rate of such objects of
${\sim}10^{-4}$). 

\begin{figure}
\begin{center}
\includegraphics[width=\columnwidth]{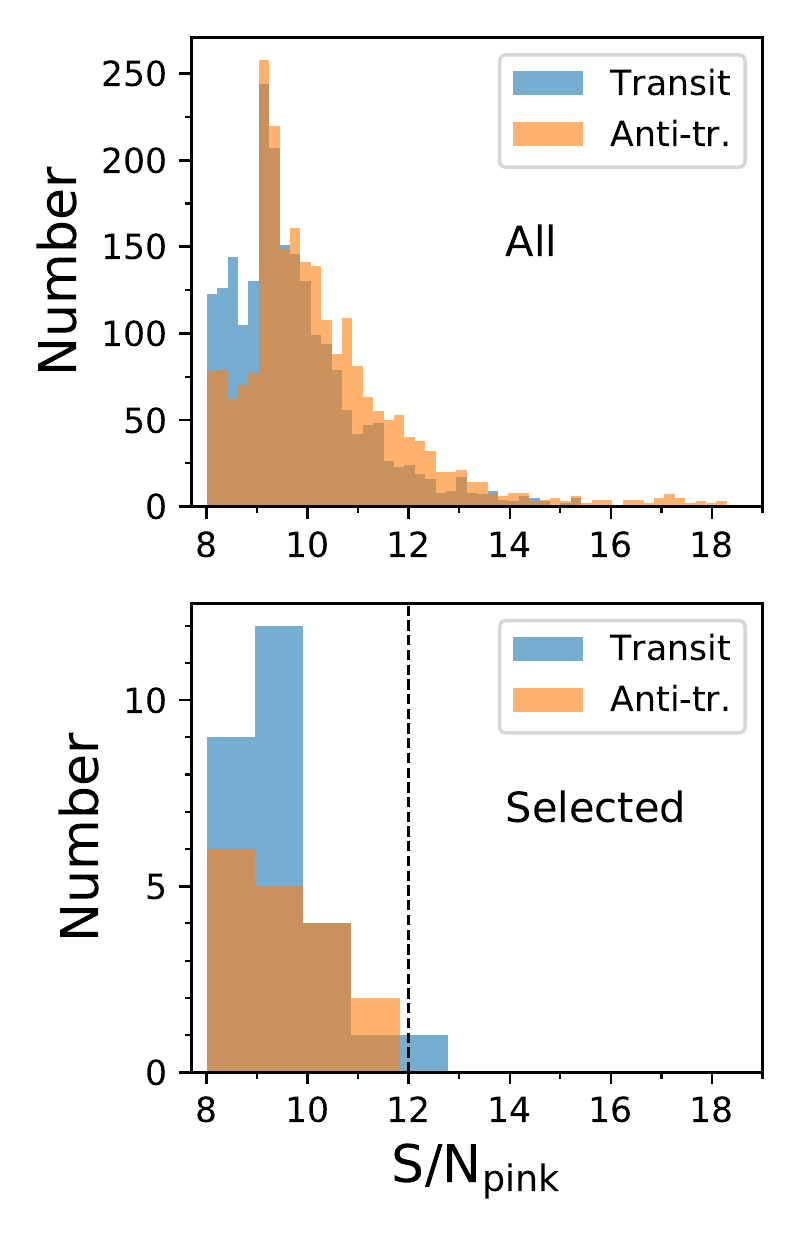}
\end{center}
\caption{\label{fig3:spn} Histograms of signal-to-pink noise (S/PN)
  for all signals that exceed our thresholds and for our selected
  planet candidates.  The distributions for the transits are shown
  in blue and the distributions for the anti-transits (labeled as
  ``Anti-tr.'' in the 
  figure) are shown in orange.  The top panel shows all the values
  that cross our S/PN, transit duration, number of transits, and number
  of points in transit thresholds (see text) and the bottom panel shows just
  those candidate signals that were selected in our by-eye vetting.
  The vertical line in the
bottom panel shows our chosen S/PN cutoff value of 12.}
\end{figure}

The by-eye vetting was performed again, with both the transit and anti-transit results presented.
  The light curves
presented for the anti-transit search results were inverted so that
the signals would appear as transits. No special indication was
given during the manual vetting as to whether a given signal was a
transit or anti-transit, permitting a blind vetting of the signals.
We implemented several cuts based on the BLS statistics.  Only those
signals with S/PN${>}8$, $q/q_\mathrm{exp}\geq0.25$ (or 0.5 if
$8<\mathrm{S/PN}<9$), number of transits $n_\mathrm{t}\geq 3$, and number of
points in transit $n_\mathrm{pit}\geq 15$ were examined.
We recorded all signals that we thought were
possible transits, identifying 27 transits and
17 anti-transits as planet candidates (though here we call the
anti-transits 
``planet candidates,'' we note that since these are not transits they
cannot be actual planet candidates). 
We then examined the
distributions in S/PN for both the transits and anti-transits.
These distributions are shown in Figure~\ref{fig3:spn}.  A
Kolmogorov-Smirnov (KS) test of the distributions for our 27 transit and 17 anti-transit planet
candidates has a $p$-value of 0.51 and a KS statistic of 0.24,
indicating that we cannot reject the hypothesis that the
planet-candidate transits and anti-transits are drawn from different
distributions of S/PN.

Based on these results, it is possible our
planet candidate signals are due to correlated noise and are thus false alarms.  Despite this,
we present our strongest candidates in Section~\ref{sec3:transitsearch}.
Based on our results, we decided that a S/PN cutoff value of 12 would be used in
our transit--injection--recovery pipeline to quantify our sensitivity to planetary transits.  We do have one candidate with S/PN$>12$, W2282, but with an S/PN value of 12.3, it is still only of marginal significance.

\subsection{Occurrence Rate Calculation}

\begin{figure}
\begin{center}
\includegraphics[width=\columnwidth]{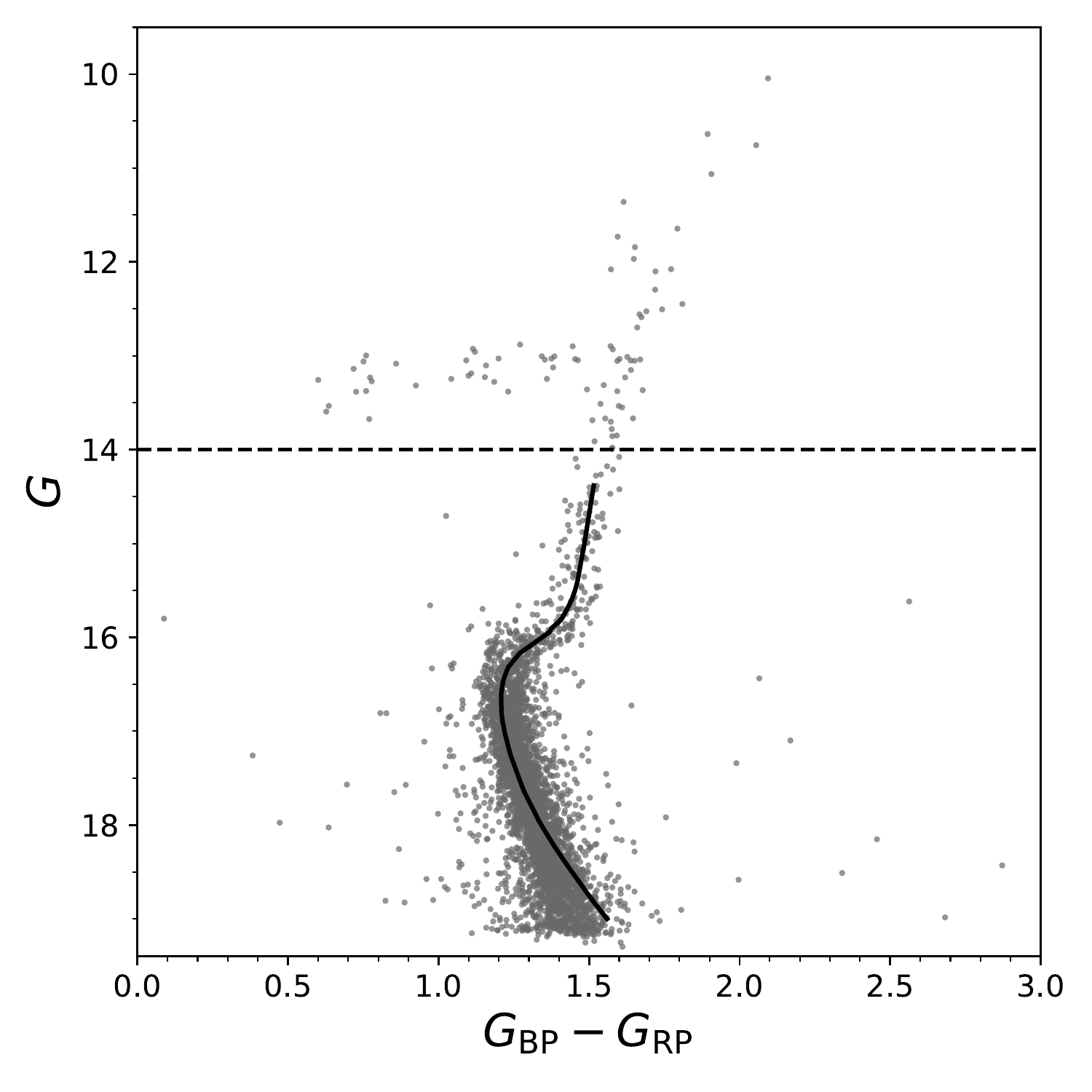}
\end{center}
\caption{\label{fig3:cmd} Color--magnitude diagram for objects in our
  analysis with a cluster membership probability
  ${>}$99\%. The photometric data are taken
  from {\it Gaia} DR2 (\citealt{gaiadr2,riello2018}).  The
  horizontal dashed line shows our magnitude cut for objects
  considered in our occurrence rate calculation, with objects below
  the line being included. The solid black line shows the isochrone 
  fit used for transit injection and recovery, as described in the text.}
\end{figure}

We focused our occurrence rate calculation only on stars that are likely cluster members by including only those objects with
membership probabilities greater than 99\% as calculated by
\cite{wallace2018} using {\it Gaia} DR2 proper motions.  There were 3784
such objects. W2282, the S/PN${>}$12 star, is among the cluster members.
For our occurrence rate calculation, we also decided to focus only on
main sequence and subgiant stars.
We imposed a cutoff of $G{>}14$ to
focus on these objects, leaving us with 3704 objects for the 
 calculation.  Figure~\ref{fig3:cmd} shows a color--magnitude diagram (CMD) of the
cluster members in our analysis with this cutoff indicated.

As a first step to calculating occurrence rates from our results, we quantified
our transit detection efficiency.  To do this, we injected transits into 
our light curves to test how well we could recover them.
The transits were injected  using \texttt{VARTOOLS}, based on the transit model of \cite{mandel2002}.
The injected periods and planet radii were taken from a 5x5 grid, with
periods drawn uniformly from uniform bins in period between one and 36
days and planet radii drawn log-uniformly from log-uniform bins
between 0.3 and 2.0 \rj.  

The stellar radii and masses were determined 
 from the PARSEC stellar evolution models
\citep{marigo2017}, obtained through the CMD v3.2
web interface\footnote{\url{http://stev.oapd.inaf.it/cgi-bin/cmd}},
this time making direct use of the {\it Gaia} DR2 $G$ magnitudes and 
using the bolometric corrections for the {\it Gaia} band-passes from
\citet{maizapellaniz2018}.  Through trial-and-error we determined the
best fit to the $G$ vs. $G_\mathrm{BP}-G_\mathrm{RP}$ CMD to be provided by a PARSEC isochrone with an age of
12.5 Gyr, a metallicity of [Fe/H]${=}-1.2$, a distance of 1.8 kpc,
extinction in the $G$ band $A_G=1.4$, and reddening
$E(G_\mathrm{BP}-G_\mathrm{RP})=0.57$. This isochrone is shown in Figure~\ref{fig3:cmd}.  The stellar parameters were
determined from the isochrone using just $G$ magnitudes. There
were seven objects with $14.0<G<14.28$ for which the isochrone interpolation as we had implemented it
failed: W364, W642, W1643, W1898, W1912, W2757, and W3684 (using the
identifiers from \citealt{wallacem4}).  We exclude these objects from the occurrence rate calculation.

Random eccentricities, phases,
longitudes of periapsis, and inclinations were chosen for each injected transit.  Phases and
longitudes of periapsis 
were chosen uniformly between zero and $2\pi$, inclinations were chosen
uniform in $\cos{i}$ subject to the constraint that transits actually
occur, and eccentricities
were drawn from a Beta distribution, with parameters as determined by the empirical fit of
\cite{kipping2013} to his short-period planets.  The planet mass was
fixed at 0.8 \mj\ for all transit injections independent of injected radius, as the simulated transit signal is effectively independent of the planetary mass.  Limb darkening was
incorporated with a quadratic model, using the parameters determined
by \cite{claret2018} for {\it Kepler} using the PHOENIX-COND model
\citep{husser2013}.

The transit-injected light curves were then ran through the same
photometric processing as the light curves searched for planetary
transits: decorrelation of systematic brightness variations against
the telescope roll and TFA.  Due to time constraints, we were unable
to run a full BLS search for each transit-injected light curve.  Instead, we used the
\texttt{-BLSFixPer} option of \texttt{VARTOOLS} to perform a BLS
search at only the injected period in order to get the BLS statistics.
Our S/PN cut of 12 was applied, as well as the additional
cuts used in our planet search (see Section~\ref{sec:spnthreshold}), namely: $q/q_\mathrm{exp}>0.25$,
$n_\mathrm{t}\geq3$, and 
 $n_\mathrm{pit}\geq15$.
Using \texttt{-BLSFixPer} is effectively a conservative approach, as
targets that have S/PN${<}12$ at the injected frequency, but
S/PN${>}12$ at other frequencies (such as a harmonic of the transit
period), will be excluded in our search whereas they may have been
recovered in a full BLS search. 

We ran some initial reconnaissance runs of our
transit--injection--recovery pipeline with a coarser period--radius
grid consisting of three period bins and four fixed planet radii and
12 samples from each bin.  The
periods were not sampled uniformly from each period bin, but rather
from a range of 
the smallest periods in each bin.  The purpose of these runs was to
determine, star-by-star, parameter ranges in which we might expect
to have a near-0\% transit recovery rate.  This information could then be used to accelerate the subsequent calculations.  Applying just the S/PN cut
and not the other cuts in $q/q_\mathrm{exp}$, $n_\mathrm{t}$, and $n_\mathrm{pit}$, we found 442 objects for which none of the 144
injected transits were recovered.  Many of these objects 
were significantly blended with brighter and/or variable objects and
their light curves
had very large scatter. These objects were removed from
subsequent consideration, which, with the seven objects that were not
fit by the isochrone, left us with 3255 objects that were included in
our final transit--injection--recovery analysis.  Additionally, for a given object, those
period bin/planet radius pairs  that had no recovered transits
were recorded. Injected transits with periods equal to or longer 
 and planet radii equal to or smaller than the values represented by these period bin/planet radius pairs  were
automatically recorded as  
non-recoveries.  This was about 50\% of our injected transits.
If this approximation lead
to us missing some injected transits that may have been recovered,
then our final occurrence rate upper limits will be higher than we would
have otherwise calculated, making this a conservative approximation
for the upper limits.

We then injected 56 transits into each of the raw light curves for each
radius--period bin in our $5\times5$ grid and ran them through our photometric processing pipeline.  Recovered transits were then determined based on the BLS statistics and associated cuts as discussed.
 Then for each radius--period bin, we calculated the number, $N$, of
expected planets that we would have detected if every star hosted one
planet in that bin, using (from, e.g., \citealt{ford2008})
\begin{equation}
\label{eq:n}
N = \sum_i^{n_*}\frac{1}{n_{i}}\sum_j^{n_{i}}   \delta_{ij} \frac{(\mathrm{R}_{*,i} +
  \mathrm{R}_{p,ij})(1 - e_{ij}\cos{\varpi_{ij}})}{a_{ij}\times(1-e_{ij}^2)},
\end{equation}
where $i$ is an index over the stars examined, $n_*$ is the number of
stars examined, $j$ is an index over the
individual transit injections, 
 $n_i$ is the number of
transit injections performed for the star,
$\delta_{ij}$ is one if the particular
injected transit is recovered and zero if not, R$_{*,i}$ is the
stellar radius of the star $i$ (based on the isochrone interpolation; this is the
same stellar radius used for the transit injection), R$_{p,ij}$ is the
radius of the planet for the given injected transit (taken as the
actual value used for the transit injection rather than a calculated radius
recovered from the transit signal), and
$e_{ij}$, $\varpi_{ij}$,
and $a_{ij}$ are respectively the eccentricity, longitude of
periapsis, and semimajor axis of
the orbit of the injected transit.  The quantity 
$(\mathrm{R}_{*,i} + \mathrm{R}_{p,ij})(1 - e_{ij}\cos{\varpi_{ij}})/[a_{ij}\times(1-e_{ij}^2)]$
accounts for the probability of
  transit given the random inclinations of orbits.

Once $N$ is calculated for a given radius--period bin, the 3$\sigma$,
99.73\% confidence interval upper limit for the occurrence rate assuming
no detections is calculated using the binomial distribution, with $N$
rounded to the nearest integer.  For the bin in which our S/PN$>12$ planet
candidate falls, the 3$\sigma$ confidence interval for the occurrence
rate is also calculated.  When calculating occurrence rates and limits for comparisons with other works, Equation~(\ref{eq:n}) is again used to
calculate the expected number of planets but with injected transits
chosen from a selected radius--period range instead of just the fixed
bins we drew from for the transit injections.

In performing the calculation as we have, there is an implicit
assumption that 100\% of our injected transits would appear in the five
highest BLS peaks in the full BLS search (since that is the number of peaks use din our planet search).  There is also an assumption
that 100\% of injected transits that exceed our
cutoff values would be identified in our by-eye analysis.  We ran a
full BLS calculation on a subset of our injected transits
and found that 97.5\% of injected transits that exceed our cutoff
values appear in the five highest BLS peaks.  We also performed a
by-eye vetting of approximately 500 injected transits that exceed our
cutoff thresholds with
$12<\textrm{S/PN}<12.1$ and found 98.8\% passed our vetting.  Presumably an
even larger fraction of those with higher S/PN values would pass the
by-eye vetting.  Based on these results, we decided to maintain our
assumptions of 100\% recovery for both of these steps.

\section{Transit Search Results}
\label{sec3:transitsearch}

\begin{figure*}
\begin{center}
\includegraphics[width=\textwidth]{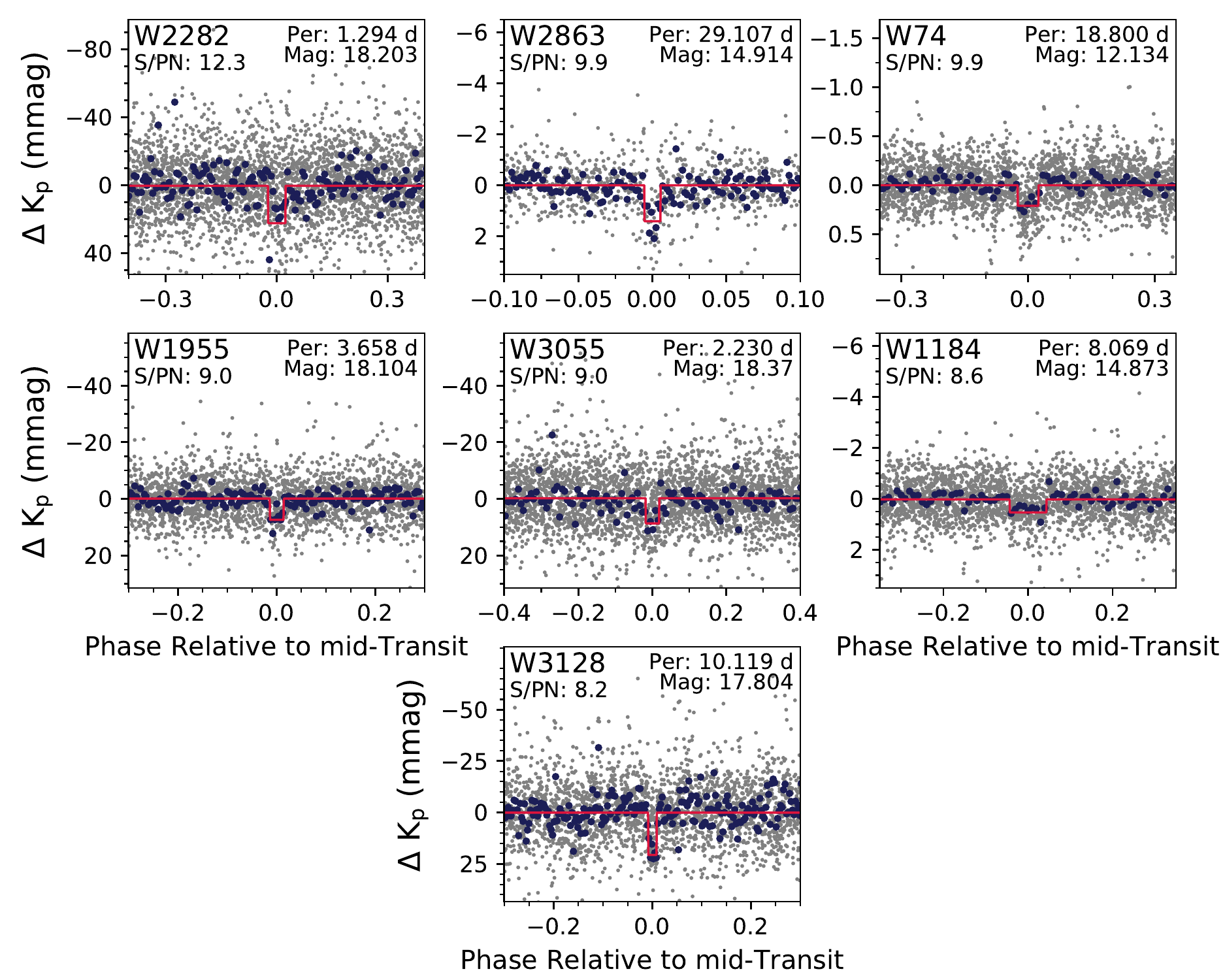}
\end{center}
\caption{\label{fig3:plancands}Phase-folded light curves for the
  transit signals of the best planet candidates.  Each panel is
for a different candidate.  In the upper-left corner of each panel is
shown, from top to bottom, the object's identifier and the 
S/PN.  In the upper-right corner is shown, from top to bottom,
the period in days and the median magnitude subtracted off for the
light curve.  In each panel, the gray points are the individual
measurements (subject to a 5$\sigma$ sigma clipping with three iterations)
and the blue points are binned-weighted-mean values.  The red line shows the
BLS fit to each phase-folded light curve.}
\end{figure*}

Figure~\ref{fig3:plancands} shows the phase-folded light curves of our seven most promising 
planet candidates. We choose not to present the other 20 candidates
that initially passed our by-eye vetting as we now consider these to
almost certainly be false alarms.  Table~\ref{tab:plan_stars} presents
information on the 
stars hosting these planet candidates and Table~\ref{tab:plan_planets}
presents information for each of the transit signals and calculated
planet properties.  The uncertainties on the periods and times
  of transit center were calculated using a Markov Chain Monte Carlo
  fit of the transit models of \cite{mandel2002} to the data.
All of the planet candidates presented are proper motion members of the cluster (\citealt{wallace2018}).  Except for W2282, all of these objects fall below our S/PN threshold
of 12, and W74 is brighter than our $G$ threshold of 14. Owing to 
the potential scientific impact of discovering a transiting exoplanet in a GC, we
choose to present the most promising candidates we found irrespective of these
cuts.  That being said, the relatively low S/PN values these signals
have indicates that most, if not all, of these candidates are likely
false alarms. 
The phase-folded light curves of 
W74, W1184, W2863, and W3128 appear to be the most robust of the
seven, while the other three appear less robust.  
Given the 
high probability that these signals are false alarms, 
follow up is needed before they are confirmed.  The next step in
following these up would be to confirm the transits and then look for
background objects to ensure these objects are not blended eclipsing
binaries.  For \simm 1-\rj\ objects, additional RV follow
up would be needed to measure the masses to identify them as planets,
brown dwarfs, or late M-dwarfs.  For 
Neptune-sized objects and smaller, the follow-up photometric data may be
sufficient to classify the objects as planets without RV
data. This is because there are no known astrophysical objects with
radii comparable to Neptune orbiting stars, so if the data permit a
sufficiently precise measurement of the radius of a Neptune-sized
object and are able to rule out blended eclipsing binaries (e.g., by showing
a lack of secondary eclipses and ellipsoidal variability), direct confirmation
from the transit data alone is possible.

\begin{deluxetable*}{ccccccc}
\tablewidth{0pc}
\tablecolumns{7}
\tablecaption{Information on Stars Hosting Planet Candidates\label{tab:plan_stars}}
\tablehead{
\colhead{ID\tablenotemark{a}} & \colhead{{\it Gaia} DR2 ID\tablenotemark{b}} & \colhead{R.A.\tablenotemark{c}} & \colhead{Decl.\tablenotemark{c}} & \colhead{$G$\tablenotemark{d}} & \colhead{Radius\tablenotemark{e}} & \colhead{Mass\tablenotemark{f}} \\
 \colhead{} & \colhead{} & \colhead{(hh:mm:ss)} & \colhead{(dd:mm:ss)} & \colhead{(mag)} & \colhead{(R$_\sun$)} & \colhead{(M$_\sun$)}  }
\startdata
W2282 & 6045466502667197056 & 16:23:34.95 & $-$26:29:14.2 & 18.32 & 0.68 & 0.66\\
W2863 & 6045466640106160128 & 16:23:41.10 & $-$26:28:04.2 & 15.07 & 3.4 & 0.80\\
W74 & 6045477635223138432 & 16:22:57.99 & $-$26:28:46.8 & 12.30 & 13 & 0.86\\
W1955 & 6045503091478311808 & 16:23:31.71 & $-$26:22:33.7 & 18.23 & 0.69 & 0.67\\
W3055 & 6045501996278191104 & 16:23:43.33 & $-$26:25:06.1 & 18.57 & 0.64 & 0.64\\
W1184 & 6045478597295755520 & 16:23:22.89 & $-$26:27:04.2 & 15.02 & 3.5 & 0.80\\
W3128 & 6045466429642662272 & 16:23:44.26 & $-$26:29:12.0 & 18.13 & 0.71 & 0.68\\
\enddata
\tablecomments{All of these stars are proper motion cluster members \citep{wallace2018}.}
\tablenotetext{a}{The identifier by which the object is known in this work, the same as in \cite{wallacem4}.}
\tablenotetext{b}{{\it Gaia} DR2 source ID.}
\tablenotetext{c}{J2000.0; data taken from  {\it Gaia} DR2 \citep{lindegren2018}.}
\tablenotetext{d}{{\it Gaia} $G$ magnitude from {\it Gaia} DR2 \citep{riello2018}.}
\tablenotetext{e}{The radius of the star in units of solar radii, determined from an isochrone fit.}
\tablenotetext{f}{The mass of the star in units of solar mass, determined from an isochrone fit.}
\end{deluxetable*}

\begin{deluxetable*}{cccccccc}
\tablewidth{0pc}
\tablecolumns{8}
\tablecaption{Information on Planet Candidates\label{tab:plan_planets}}
\tablehead{
\colhead{ID\tablenotemark{a}} & \colhead{Period\tablenotemark{b}} & \colhead{T$_0$\tablenotemark{c}} & \colhead{Depth\tablenotemark{d}} & \colhead{Radius\tablenotemark{e}} & \colhead{$q$\tablenotemark{f}} & \colhead{$q/q_\mathrm{exp}$\tablenotemark{g}} & \colhead{S/PN\tablenotemark{h}} \\
 \colhead{} & \colhead{(day)} & \colhead{(KBJD)} & \colhead{(mmag)} & \colhead{(R$_\mathrm{J}$)} & \colhead{} & \colhead{} & \colhead{}  }
\startdata
W2282 & $1.2937\pm0.0014$ & $2061.955\pm0.040$ & 22 & 1.0 & 0.047 & 0.82 & 12.3\\
W2863 & $29.07\pm0.86$ & $2078.65\pm0.67$ & 1.4 & 1.3 & 0.011 & 0.36 & 9.9\\
W74 & $18.804\pm0.057$ & $2071.675\pm0.077$ & 0.21 & 1.9 & 0.049 & 0.33 & 9.9\\
W1955 & $3.6608\pm0.0020$ & $2062.6319\pm0.0068$ & 7.5 & 0.62 & 0.028 & 1.0 & 9.0\\
W3055 & $2.20858\pm0.00054$ & $2062.1311\pm0.0063$ & 8.9 & 0.62 & 0.037 & 1.0 & 9.0\\
W1184 & $8.08\pm0.10$ & $2066.84\pm0.28$ & 0.51 & 0.82 & 0.087 & 1.2 & 8.6\\
W3128 & $10.118\pm0.090$ & $2063.465\pm0.094$ & 21 & 1.1 & 0.017 & 1.1 & 8.2\\
\enddata
\tablecomments{All of these stars are proper motion cluster members \citep{wallace2018}.}
\tablenotetext{a}{The identifier by which the object is known in this work, the same as in \cite{wallacem4}.}
\tablenotetext{b}{The period of the transit signal in days.}
\tablenotetext{c}{The time of transit center, in KBJD.}
\tablenotetext{d}{The depth of the transit signal in millimagnitudes.}
\tablenotetext{e}{The calculated radius of the planet in Jupiter radii based on the transit depth and isochrone-based stellar radius.}
\tablenotetext{f}{The fractional transit duration.}
\tablenotetext{g}{The ratio of the fractional transit duration with the expected $b=0$ transit duration.}
\tablenotetext{h}{The signal-to-pink noise of the signal.}
 \end{deluxetable*}

\clearpage

\begin{figure}
\begin{center}
\includegraphics[width=\columnwidth]{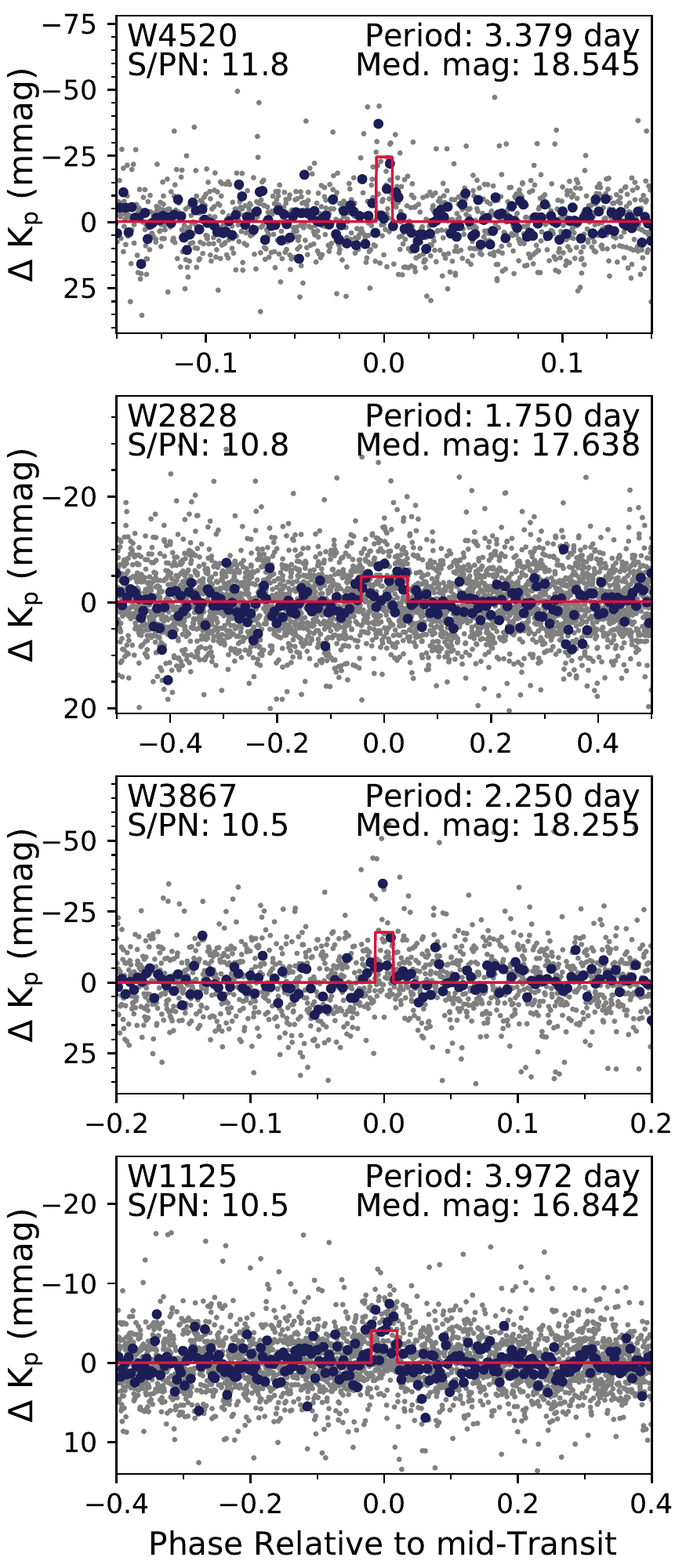}
\end{center}
\caption{\label{fig3:inversecands}Same as Figure~\ref{fig3:plancands}
  but for a few representative anti-transits.  These are presented as
  examples of the false alarms that can be produced by the systematic noise that exists in our data.}
\end{figure}

\begin{figure}
\begin{center}
\includegraphics[width=\columnwidth]{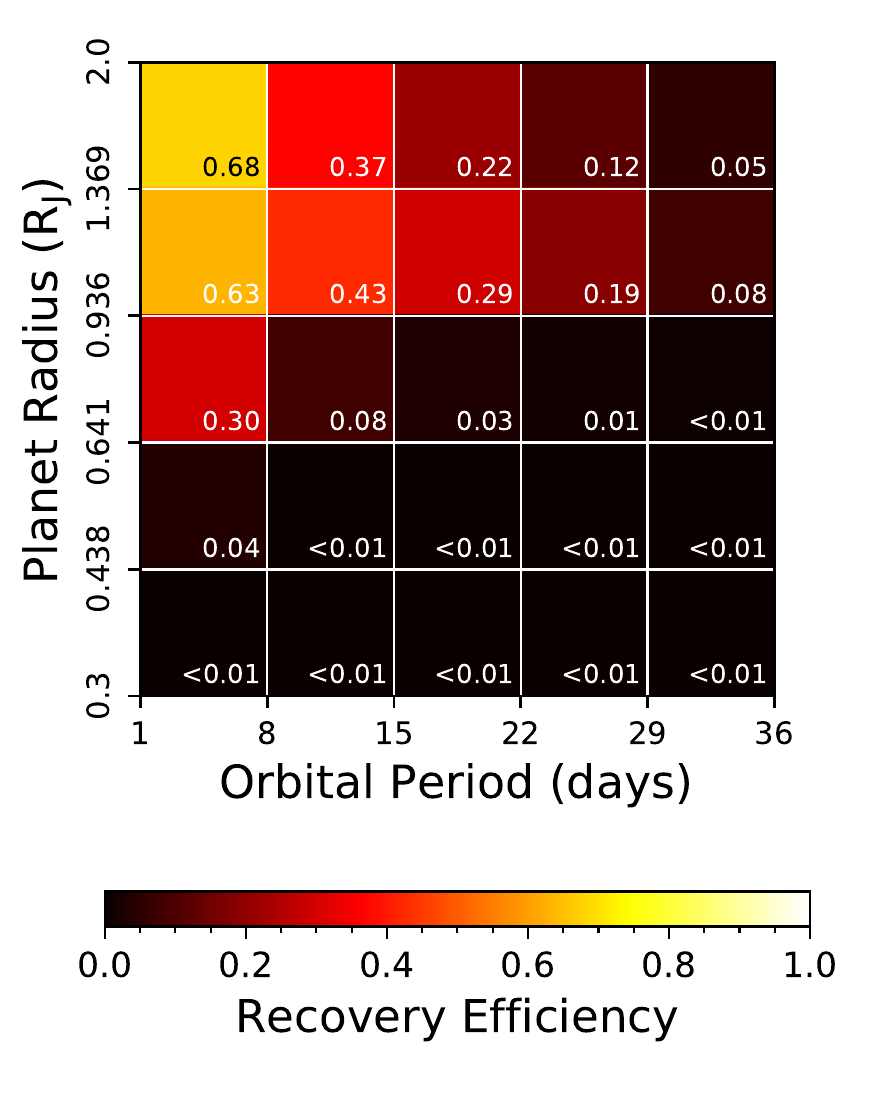}
\end{center}
\caption{\label{fig3:efficiency}Recovery efficiency of our
  transit--injection--recovery pipeline.  Each bin shows the fraction
  of injected transits that were successfully recovered across all the
stars.  The number in the lower-right corner of each bin shows the
efficiency value, which is also represented by the color of the
bin and the associated color bar.  In cases where the recovery
efficiency was less than 1\%, an upper limit of 1\% is shown.}
\end{figure}

Owing to the crowded nature of the cluster and of the K2 observations
in particular, blending is a virtually unavoidable aspect of the
data.  We confirmed
that W1184, W1955, W2282, W3055, and W3128 had the largest
signal amplitudes of all nearby objects at the respective transit periods,
for those objects for which we had light curves.
The results for W74 and W2863 were more ambiguous, likely owing to
their brighter magnitudes causing these stars to impact larger
areas of the images than the other fainter stars.  However, these two stars are each the
brightest stars in their areas of the images.

\begin{figure*}
\begin{center}
\includegraphics[width=5.9in]{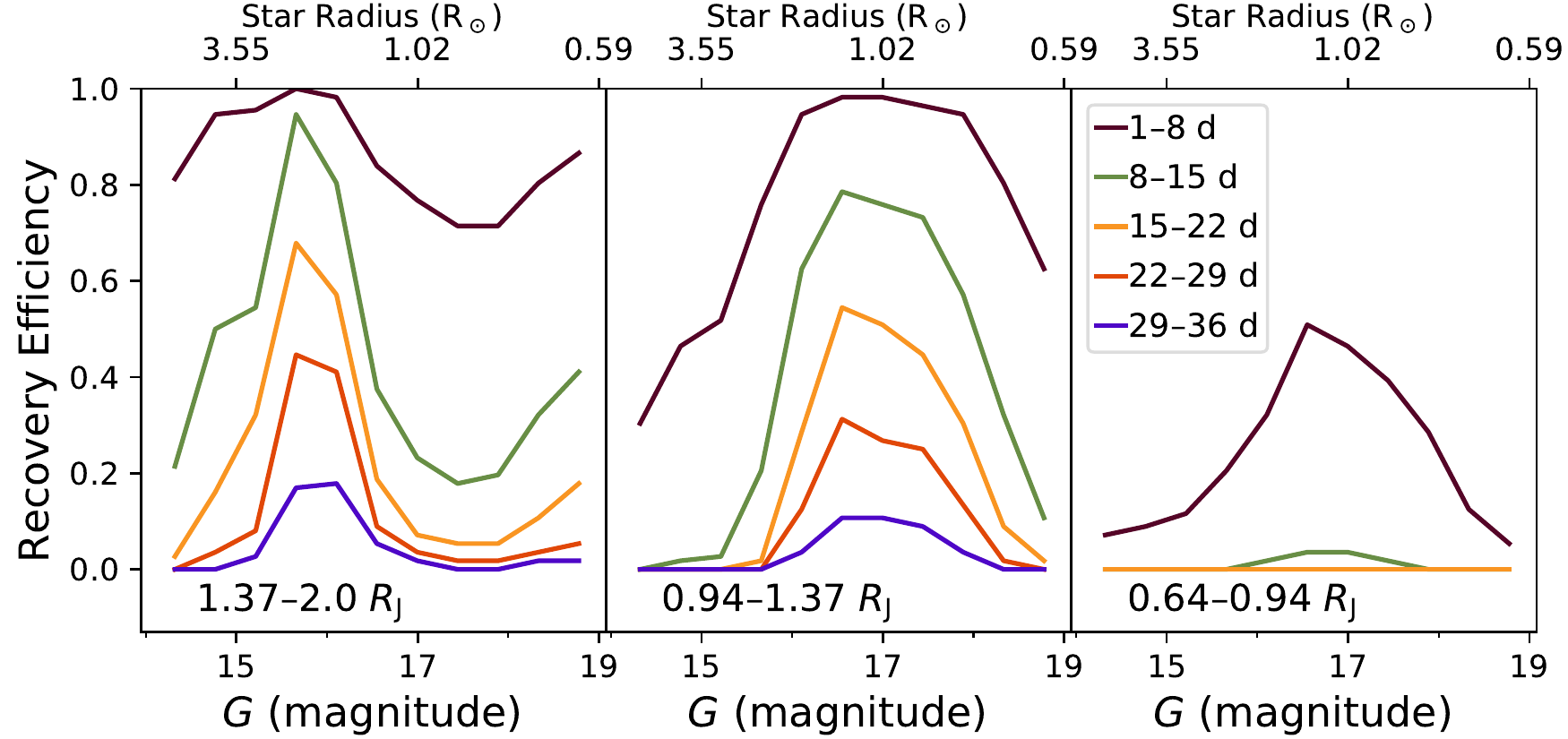}
\end{center}
\caption{\label{fig3:efficiencybyg}Recovery efficiency of our
  transit--injection--recovery pipeline, broken down by stellar
  magnitude and bins of injected planet radius and orbital
  period. Magnitude is represented along the horizontal axis, planet
  radius by the three panels, and orbital period by the color. 
  The lines show the median recovery efficiency as a function of
  magnitude across all stars for a given radius--period bin.
  In the rightmost panel, the 22--29 day (red) and 29--36 day (blue) lines
  fall behind 
  the 29--36 day (yellow) line and thus do not appear.
  The planet radius range for each
  panel is indicated in the bottom of the 
  panel.   The legend in the rightmost panel shows the color
  representation of the orbital period bins (``d'' in the legend
  stands for ``day'') and applies to all three panels. 
$G$ converted to stellar radius via an isochrone fit is shown on
  the top of each panel.}
\end{figure*}

W74 merits some additional comments.  Its CMD
position puts it on the 
red giant branch and it is asteroseismically active.  A Generalized Lomb--Scargle (GLS, \citealt{lomb1976,scargle1982})
search reveals significant sinusoidal variability at a variety of
periods (though not the \simm 19 days found by BLS), with the strongest variability at
0.77 and 1.69 days.  Pre-whitening the light curve by running LS three
times and removing a two-harmonic and one-subharmonic fit to the peak
period each time prior to running BLS recovers a similar
period as before---18.843 days---and a comparable though lesser S/PN value of 9.6.
It also bears mentioning that the \simm 18.8 day period of this
object, based on our \cite{yi2001} isochrone fit for the mass and radius
of the star, has a semimajor axis of \simm 28 R$_\sun$, compared to
the calculated stellar radius of \simm 13 R$_\sun$.  This is a
physically plausible scenario, but again this is a blended object and
we were not able to conclusively determine if the transit belonged to
this object.  A blended eclipsing binary scenario is also possible.
The implied planet
radius based on the calculated stellar radius is 1.9 \rj.

A few of our ``planet candidate'' anti-transits are shown in Figure~\ref{fig3:inversecands}
as examples of the kinds of false alarms the correlated noise in our light curves can produce.  While the S/PN values are comparable to those of our
prospective planet candidates, we think that a qualitative, by-eye
evaluation of the signals show W74, W1184, W2863, and W3128 in particular to
be more physical and transit-like than even the highest S/PN
anti-transits.  Also, those four objects have much longer
periods than any of the anti-transits we identified, suggesting that
the transit-mimicking correlated noise may exist only at shorter
periods and that these longer-period signals may be more likely to be real.

\section{Occurrence Rate Results}
\label{sec3:occurrence}

\subsection{Transit Recovery Results}

\begin{figure}
\begin{center}
\includegraphics[width=\columnwidth]{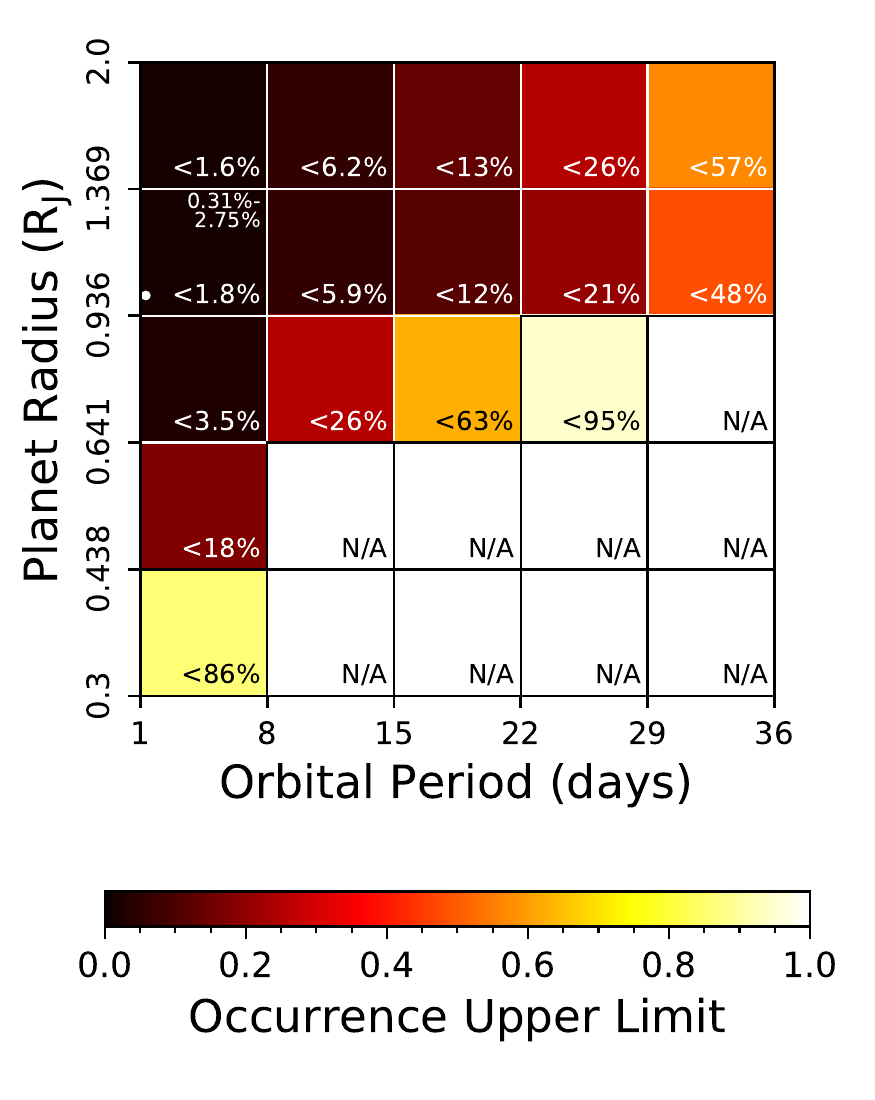}
\end{center}
\caption{\label{fig3:occurrence}Calculated upper limits on occurrence
  rates for our radius--period bins.  The lower-right hand of each bin
shows the 3$\sigma$ upper limit in the fraction of stars having at
least one planet in that bin as calculated using a binomial
distribution based on our determined
detection efficiencies and transit probabilities. Those bins marked
``N/A'' either had too low of detection efficiencies for us to
calculate any occurrence rate or had a rate that was indistinguishable
from 100\%. The color of each
bin is a representation of the occurrence rate upper limits, based on
the color bar at the bottom of the figure.  The white point
represents the one planet candidate we found that passes our S/PN
threshold, W2282, and the range in the upper-right corner of the
associated bin is the 3$\sigma$ range on the occurrence rate assuming
the planet candidate is real.}
\end{figure}

Figure~\ref{fig3:efficiency} shows the recovery efficiency of our
injected transits  across our radius and period bins. We define recovery
efficiency as the fraction of transits 
that were successfully recovered, and in Figure~\ref{fig3:efficiency}, this is the efficiency across all of the injected transits and all of the stars. As would be expected, the recovery efficiency
trends towards higher values for larger planets and
smaller orbital periods. 
Of note, though, is that for 
period bins greater than 8 days, the recovery efficiency is higher for
our second-largest planet radius bin (\simm 0.9 to \simm1.4 \rj)
than for our largest planet radius bin (\simm1.4--2 \rj).  A possible
explanation for this is that the deeper transits produced by the
larger radius planets were more likely to be distorted and diminished
by our photometric processing pipeline than the shallower transits
from 0.9--1.4 \rj\ planets.  The shorter-duration transits at smaller
periods may be less likely to be impacted by the processing pipeline,
which would explain the higher recovery efficiency seen for the larger
planets for periods shorter than eight
days. Also, we found that some of the deepest transits had the bottom
portions of the transits trimmed by the sigma clipping.  Such transits
were still detectable by BLS but had a lower S/PN due to the diminished
apparent transit depth.

Figure~\ref{fig3:efficiencybyg} shows the recovery efficiency broken
down by $G$ magnitude, orbital period, and planet radius.  
We see (as
expected) that shorter-period planets have a higher recovery
efficiency than longer-period planets. We also see lower recovery
efficiencies for the brightest stars relative to the peak efficiencies
reached 
(usually around $G{\approx}16$--17).  This is due to the large radii of
the brightest stars (see top axis of Figure~\ref{fig3:efficiencybyg}) diminishing the transit depth and thus
the signal size and recoverability.
In the leftmost panel of Figure~\ref{fig3:efficiencybyg},
corresponding to the largest-radius injected planets, we also note that
 detection efficiencies tend to
be higher at both $G{=}16$ and $G{=}19$ than at
$G{=}18$, particularly in the 1--15 day period range.
The non-monotonic variation in detection efficiency with magnitude is
due to the different magnitude dependencies of two competing
effects. Brighter stars in the cluster have higher precision light
curves, which tend to increase  
the signal to noise of the transits. However, fainter stars in the
cluster have smaller stellar radii, which leads to deeper transits for
a given planetary radius. 
In the middle and rightmost panels of Figure~\ref{fig3:efficiencybyg}, 
this increase in
recovery efficiency at the faintest magnitudes  is not seen.  
In the rightmost panel,  we see a large drop in recovery
efficiency overall for 0.64--0.94 \rj\ planets relative to the other two panels---the larger radii planets.

\subsection{Planet Occurrence Rates and Limits}

We now present our calculated occurrence rate limits and compare with
other published occurrence rates.  Figure~\ref{fig3:occurrence} shows
the calculated occurrence rate upper limits across our radius--period
bins, and in the case of the bin containing our single S/PN$>12$ planet
candidate  (W2282), the range for the
occurrence rate if the planet candidate is real.  For our
shortest period bins, we are able to get down to limits of 1.6--3.5\% for
bins with planet radius larger than 0.64 \rj.  To put these limits in context,
Table~\ref{tab:comp1} compares our occurrence rate limits with those of
works using {\it Kepler}, {\it TESS}, or RV surveys for field stars. These previous works are: {\it
  Kepler}-based occurrence rates from \cite{howard2012},
\cite{fressin2013}, \cite{masuda2017}, and \cite{petigura2018}; a {\it
  TESS}-based occurrence rate from \cite{zhou2019}; and RV occurrence
rates from \cite{mayor2011} and \cite{wright2012}.
Table~\ref{tab:comp2} compares our occurrence rate limits with those
of previous GC planet searches: \cite{gilliland2000} and \cite{weldrake2005}
for 47 Tuc, \cite{weldrake2008}
for $\omega$ Cen, and \cite{nascimbeni2012} for NGC 6397.

\begin{deluxetable*}{cccccc}
\tablewidth{0pc}
\tablecolumns{6}
\tablecaption{Comparison with {\it Kepler}, {\it TESS}, and RV
  Occurrence Rates for Field Stars\label{tab:comp1}}
\tablehead{
  \colhead{Per. Range\tablenotemark{a}} & \colhead{Rad. Range\tablenotemark{b}} & \colhead{Reference}& \colhead{Published Rate\tablenotemark{c}}  &\colhead{Our Upper Limit\tablenotemark{d}} & \colhead{Our Rate\tablenotemark{e}}\\
 \colhead{(day)} & \colhead{(\rj)} & \colhead{} & \colhead{} &\colhead{} & \colhead{}  }
\startdata
\cutinhead{{\it Kepler} Studies}
0.8--10 & 0.71--2.85 &\cite{howard2012}& $0.4\pm0.1$\% & $<2.2$\% & 0.38--3.3\% \\
0.8--50\tablenotemark{f} & 0.71--2.85 &\cite{howard2012}& $1.3\pm0.2$\% & $<6.1$\% & 1.1--9.1\% \\
0.8--10 & 0.36--0.71 &\cite{howard2012}& $0.5\pm0.1$\% & $<16$\% & \textellipsis \\
 0.8--10 & 0.54--1.96 &\cite{fressin2013}& $0.43\pm0.05$\% & $<2.6$\% & 0.44--3.9\% \\
 0.8--17 & 0.54--1.96 &\cite{fressin2013}& $0.70\pm0.08$\% & $<3.8$\% & 0.66--5.7\% \\
 0.8--29 & 0.54--1.96 &\cite{fressin2013}& $0.93\pm0.10$\% & $<6.0$\% & 1.0--8.9\% \\
0--10 & 0.8--2 & \cite{masuda2017} &  $0.43^{+0.07}_{-0.06}$\% & $<2.1$\% & 0.36--3.2\%\\
0--10 & 0.8--2 & \cite{masuda2017} & $0.24^{+0.10}_{-0.09}$\%\tablenotemark{g} & $<2.1$\% & 0.36--3.2\%\\
1--10 & 0.71--2.14 & \cite{petigura2018} & $0.57^{+0.14}_{-0.12}$\% &
$<2.2$\% & 0.38--3.3\%\\
\cutinhead{{\it TESS} Study}
0.9--10 & 0.8--2.5 &\cite{zhou2019} & $0.41\pm0.10$\%  & $<2.1$\% &0.36--3.2\% \\
\cutinhead{RV Studies}
0--11 & 0.72--2\tablenotemark{h} & \cite{mayor2011} & $0.89\pm0.36$\% & $<2.4$\% & 0.40--3.5\%\\
0--10 & 0.55--2\tablenotemark{h} & \cite{wright2012} & $1.20\pm0.38$\% & $<2.5$\% & 0.44--3.8\%\\
\enddata
\tablenotetext{a}{The period range used in the comparison work for the occurrence rate calculation.  Note that the smallest period used in this work is 1 day and so our calculation truncates smaller period ranges at 1 day.}
\tablenotetext{b}{The planet radius range used in the comparison work for the occurrence rate calculation.  Several references used R$_\earth$ as their unit of radius and these values have been converted to \rj\ and rounded. Note that the largest radius examined in this work is 2 \rj\ and so our calculation truncates larger radius ranges at 2 \rj.}
\tablenotetext{c}{The comparison work's planet occurrence rate  as published.}
\tablenotetext{d}{Our calculated occurrence rate upper limit for the same period and radius range.}
\tablenotetext{e}{Our calculated occurrence rate assuming W2282 is a planet.  This value is not included  if W2282 does not fall in the given period and radius ranges.}
\tablenotetext{f}{Our calculation truncates at 36 days.}
\tablenotetext{g}{For this value, \cite{masuda2017} restricted their analysis to the {\it Kepler} stars that were in the same range of masses as the stars search in 47 Tuc for planets by \cite{gilliland2000}.}
\tablenotetext{h}{These were limits in mass rather than radius.  We converted the lower mass limit to a radius using the empirical relation derived by \cite{chen2017} and imposed our default upper limit of 2 \rj.}
\end{deluxetable*}

 \begin{deluxetable*}{cccccc}
\tablewidth{0pc}
\tablecolumns{6}
\tablecaption{Comparison with Globular Cluster Occurrence Rates\label{tab:comp2}}
\tablehead{
  \colhead{Per. Range\tablenotemark{a}} & \colhead{Rad. Range\tablenotemark{b}} & \colhead{Reference}& \colhead{Published Rate\tablenotemark{c}}  &\colhead{Our Upper Limit\tablenotemark{d}} & \colhead{Our Rate\tablenotemark{e}}\\
 \colhead{(day)} & \colhead{(\rj)} & \colhead{} & \colhead{} &\colhead{} & \colhead{}  }
\startdata
1--8 & 0.64--2 & \cite{gilliland2000}\tablenotemark{f} & $\lesssim 0.7$\% & $<2.1$\% & 0.35--3.1\% \\
1--16 & 1--2 & \cite{weldrake2005} & \textellipsis & $<2.7$\% & $0.47-4.1$\%\\
1--5\tablenotemark{g} & 1.5--2\tablenotemark{g} & \cite{weldrake2008} & $<1.7$\% & $<0.81$\%\tablenotemark{h} & \textellipsis\\
1--14\tablenotemark{i} & 0.94--1.37\tablenotemark{i} & \cite{nascimbeni2012} & $<9.1$\% & $<0.93$\%\tablenotemark{h} & 0.31--1.7\% \\
\enddata
\tablenotetext{a}{The period range used in the comparison work for the occurrence rate calculation.}
\tablenotetext{b}{The planet radius range used in the comparison work for the occurrence rate calculation.  The radius ranges used were not always clear in the comparison works, so we made our best guess, taking into account our radius--period grid boundaries.  Note that the largest radius examined in this work is 2 \rj\ and so our calculation truncates larger radius ranges at 2 \rj.}
\tablenotetext{c}{The comparison work's planet occurrence rate upper limit as published.}
\tablenotetext{d}{Our calculated occurrence rate upper limit for the same period and radius range.}
\tablenotetext{e}{Our calculated occurrence rate assuming W2282 is a planet.  This value is not included if W2282 does not fall in the given period and radius ranges.}
\tablenotetext{f}{Provided courtesy K. Masuda (private communication) based on the work in \cite{masuda2017}.  The calculated 3$\sigma$ occurrence rate upper limit is based on the same stellar mass range described in Table~\ref{tab:comp1}, note g.}
\tablenotetext{g}{The published rate is for their 1--5 day calculation; we performed our calculation over 1--8 days to guarantee at least one of our radius--period bins be included.  Similarly, the quoted rate is only for ${>}1.5$ \rj\ objects, but we had to use 1.37--2 \rj\ objects to cover a whole bin.}
\tablenotetext{h}{95\% confidence instead of our typical 3$\sigma$, to match the confidence level used by both \cite{weldrake2008} and \cite{nascimbeni2012}.}
\tablenotetext{i}{The upper end of the period range was arrived at dividing their total observation duration (28 days) in half; planet radius range chosen to span their single injected planet radius, 1 \rj.}
\end{deluxetable*}

As seen in Table~\ref{tab:comp1}, in no case were we able to set an
upper limit that shows an occurrence rate smaller than what we would
expect from the field population.    And even if
W2282 or a comparable planet candidate in  
the HJ regime is shown to actually be a planet, most of the
{\it Kepler}- or {\it TESS}-based occurrence rates would be
consistent with the lower end of our calculated occurrence rate ranges,
though the upper ends of our ranges are inconsistent in all those
cases.  Thus based on these previous studies, and ignoring the
metallicity dependence of HJ occurrence as seen in the field, we would
expect to 
have a non-vanishing probability of finding a planet.
Comparing with the RV studies, the RV rates fall within our
W2282-based occurrence rate ranges, also suggesting from these results
that there is some meaningful, if small, probability of
finding a planet.

Comparing with the previous searches in GCs, and focusing first on
the 47 Tuc surveys as those were the most constraining, 
\cite{masuda2017}
showed that \cite{gilliland2000} should have found  $2.2^{+1.6}_{-1.1}$ planets in
their survey, and that \cite{weldrake2005} should also have
found \simm2 planets in their survey.  Thus their results
(possibly) show a lower occurrence rate in 47 Tuc than found by {\it
  Kepler} for the field population for the period ranges searched.
Our results do not reach such a constraining level for HJs.
 However, our sensitivity reaches further in
planet radius and period than either of those two previous surveys.
Our \simm78 day baseline and the nearly continuous nature of the
observations would virtually guarantee us three visible transits for orbital
periods up to \simm26 days and for some cases out to \simm 39 days.
This is compared to the 8.3 day baseline of \cite{gilliland2000} and
the 33 day baseline of \cite{weldrake2005}.  Additionally,
\cite{gilliland2000} were insensitive to nearly all planets with a
radius below
0.8 \rj\ and had at best 40\% recovery of 1 \rj\ objects for the optimal
stellar magnitude.  Our work is still reasonably sensitive down to
\simm0.6 \rj\ for small periods and somewhat sensitive down to
\simm0.4 \rj.  

In Table~\ref{tab:comp2}, we do have
a more constraining upper limit than the work in $\omega$ Cen by \cite{weldrake2008} for their 
very limited period and radius range.
We also improve on the limit determined by \cite{nascimbeni2012}.  
They were able to put a (95\% confidence) upper limit of 9.1\% on the 
occurrence of \simm 1 \rj\ objects with periods between 1 and \simm14 days, while 
we are able to put an upper limit of 0.93\% for the same period range 
and a similar planet radius range at the same confidence level.
\cite{weldrake2005} did not provide any quantification
of their sensitivity to planet radius, but their calculations assumed
a relatively large radius of 1.3 \rj, and we assume they were
sensitive to planets of that radius and larger.  

The primary contribution of this work 
 is the new parameter range it explores
for planet occurrence rates in GCs,
in both planet radius and orbital period.  The occurrence rate limits in
these new parameter ranges ($0.3\lesssim\textrm{R}_p\lesssim0.8$
\rj\ at short periods and $P\lesssim 36$ days for large-radius
planets) are shown in Table~\ref{tab:comp1} in comparison with the
field occurrence rates.  In particular, we set occurrence rate
upper limits of 16\% for 0.36--0.71 \rj\ planets with 1--10 day periods
and 6.1\% for 0.71--2 \rj\ planets with 1--36 day periods.  While
these numbers may not seem impressive when compared to the equivalent
occurrence rates determined from {\it Kepler} ($0.5\pm0.1$\% for
0.36--0.71 \rj\ planets with 0.8--10 day periods and $1.3\pm0.2$\% for
0.71--2 \rj\ planets with 1--50 day periods; \citealt{howard2012}),
these are the first limits set for planets in a GC in these period and
radius regimes.  These limits demonstrate that the occurrence of
planets in M4 just outside the HJ regime (in terms of period or radius) is
not ubiquitous, and, for the 0.71--2 \rj, 1--36 day range, is at most
a factor of about five higher than has been found for the field population.

\section{Discussion}
\label{sec3:discussion}

This work represents the first look at a planet occurrence rate for
the GC M4, and the fifth photometry-based examination of a planet
occurrence rate for a GC.
It is worth noting that, although our results do not place very
  stringent constraints on the occurrence of planets in M4, 
{\it Kepler} was not designed or optimized for
looking at GCs---in particular, the \simm4\arcsec/pixel image resolution
led to significant blending in the images---and the superstamp
observations of M4 were originally 
intended for observing RR Lyrae variables.  Obtaining even the level of
constraints we did from a telescope and observations not originally
intended for a GC planet search is yet another demonstration of an
unanticipated scientific result from {\it Kepler}.

Though our constraints cannot rule out planet
occurrence rates for M4 matching those of the field stars, given the current uncertainties on planet
formation---particularly the formation of close-in giant planets---and
uncertainties on GC formation, obtaining any constraints on planet
occurrence in GCs for new regimes of planet radius and period is useful.
It may be that the occurrence of certain kinds of close-in planets in GCs is more
common due to some unique aspect of GCs.  

There are some reasons we might
expect the occurrence of close-in planets to be higher in a GC than in the field.  For
example, \cite{hamers2017} demonstrated that the increased number
of close stellar encounters experienced by GC stars over their
lifetimes could enhance the HJ occurrence rate for certain stellar
densities (peak formation occurred at a density of \simm$4\times10^4$ pc$^{-3}$) if 
HJs are formed through high-eccentricity migration.  The formation of
GCs themselves is still something of a mystery (see
\citealt{gratton2012} for a review), and perhaps there is something
unique about the formation of stars in GCs that would increase the formation of
close-in planets.
Our results, with those of the previous GC planet occurrence works,
provide constraints on just how enhanced a planet occurrence rate
might be should there indeed be enhanced close-in planet formation in GCs.

On the other hand, much work has been done to show
 why specifically HJ occurrence in GCs might be suppressed.
The occurrence of HJs is known to correlate with host star metallicity
(e.g., \citealt{fischer2005}) and this has
been used to argue that the low metallicities of GCs would inhibit HJ
formation (for example, \citealt{santos2001} showed the known metallicity dependence as being able to
explain the 47 Tuc planet non-detection of Gilliland et al.), but the reason
behind a metallicity--occurrence connection is not well understood
and it may be that  the underlying cause of this connection
does not apply in the unique environments of GCs. Additionally, the
dense stellar environment of GCs and the associated levels of radiation from
particularly the nearby massive stars may inhibit giant-planet
formation \citep{armitage2000,adams2004,thompson2013}. Also, in
addition to enhancing the rate of close-in planets, dynamical
interactions with passing stars can also remove planets from planetary
systems
\citep{sigurdsson1992,davies2001,bonnell2001,fregeau2006,spurzem2009}, particularly planets on wide orbits. 
Interactions between stars and protoplanetary disks lead to decreases
in disk sizes as well \citep{breslau2014}.  Until
better constraints or actual occurrence rates are determined for GCs,
for a larger range of planet radii and orbital periods than are presently accessible from existing data,
it will be difficult to determine the precise impact a GC environment has
on planet formation and occurrence.

\begin{figure}
\begin{center}
\includegraphics[width=\columnwidth]{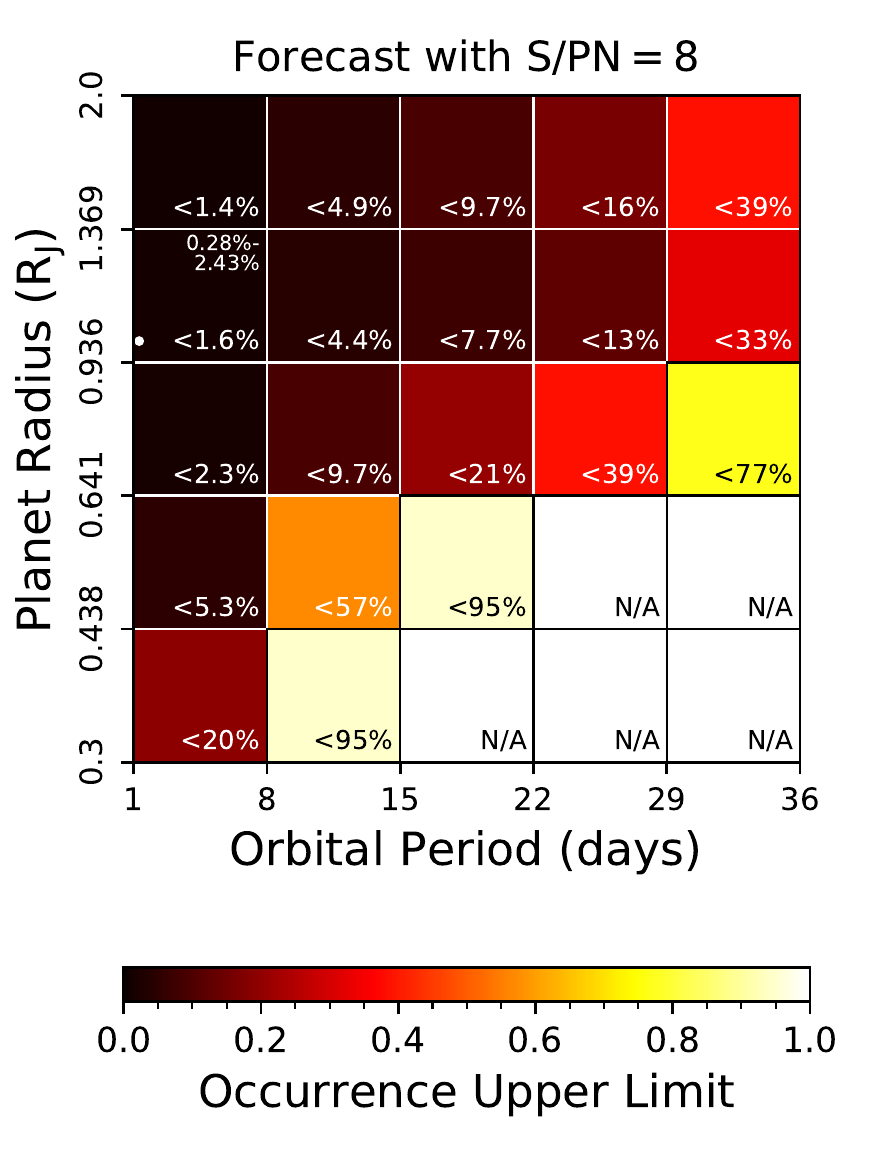}
\end{center}
\caption{\label{fig3:forecast}Same as Figure~\ref{fig3:occurrence},
but instead showing a forecast of the limits that could be set if  a S/PN threshold of eight
was able to be used instead of 12 and no additional planet candidates were found.}
\end{figure}

As an analysis of how an improvement on our light curves and/or noise
characterization and removal could improve our occurrence rate limits,
we show a forecast in Figure~\ref{fig3:forecast} of
the limits that would be set if  an 
S/PN threshold of eight could be imposed instead of 12 and assuming no
planets were found.  W2282 and the associated occurrence rate is still
included  
for comparison with Figure~\ref{fig3:occurrence}.  Our upper limits in
the HJ regime would not improve by very much, but we would be able to
place more stringent constraints for
$0.64~\mathrm{R}_\textrm{J} \lesssim \mathrm{R}_p \lesssim 0.94~\mathrm{R}_\textrm{J}$
across all the periods examined, and for
$0.3~\mathrm{R}_\textrm{J} \lesssim\mathrm{R}_p \lesssim 0.64~\mathrm{R}_\textrm{J}$ for
the shortest periods examined here.  Even if our limits in the HJ
regime do not improve much, a better understanding of the noise would
allow for an improved vetting of the current planet candidates.

As limited as our constraints are, they may be the best to come for a
while.  The only near-term continuous photometric survey is {\it
  TESS}, but with its \simm20\arcsec/pixel image resolution it will leave
most of the stars in GCs hopelessly blended.  Moreover,  the \simm1
month observation span most of its survey field will be covered by
is only about a third the span of what is available in this work with K2.  An {\it HST}
 campaign similar to that of \cite{gilliland2000}
for M4---being about half the distance as 47 Tuc---should permit a
factor of two increase in the signal-to-noise ratio for stars of comparable masses and
evolutionary state as in 47 Tuc; a campaign along these lines might be
considered. The main limiting factor in setting the HJ occurrence
limit from the K2 data is the relatively small number of cluster stars
observed, (\simm 4000 compared to \simm 34,000 GC stars in
\citealt{gilliland2000} and \simm 20,000 GC stars in
\citealt{weldrake2005}).  The K2 superstamp covered a relatively small
fraction of the stars in the cluster, so a survey that covers more of
the cluster could be useful.

Despite the low S/PN of our planet candidates, given the scientific
impact of discovering and characterizing a transiting planet in a GC,
we argue that it is worth the  effort to confirm
whether these are real planet signals.

Unfortunately, the data are already
five years old.
Referencing Table~\ref{tab:plan_planets}, the
transit timing uncertainties for W2863 (0.67 days) and W1184 (0.28
days) are considerable. Taking the number of transits
that are expected to have occurred in the last five years and
multiplying by the period uncertainties as an approximate calculation
of additional uncertainty on transit timing arising from the period
uncertainties, we get: W2282, 2.0 days;
W2863, 54 days;
W74, 5.5 days;
W1955, 1.0 days;
W3055, 0.45 days;
W1184, 23 days; and 
W3128, 16 days.  Thus only for W74, W1955, and W3055 would the transit
epoch number be known with certainty if a transit were to be observed
five years on.  In no case is the present transit timing known with
sufficient precision to be sure of catching a transit in one night's
observation, so an extended follow-up campaign would be needed.
In this 
the crowded nature of the cluster and the manageably sized field of view
of the K2 superstamp (\simm10\arcmin\ by \simm20\arcmin) are
advantages.  Many available wide field imagers can cover a large
fraction of or even the entire superstamp, allowing for simultaneous
observation of more than one planet candidate, and the data would also have the advantage of observing other interesting variables guaranteed to be present 
 (see \citealt{wallacem4} for a catalog).

\section{Conclusion}
\label{sec3:conclusion}
We searched for transiting planets in the GC M4 using data from the K2 mission. 
These
data represent the longest continuous observation of a GC,
permitting a search for the longest-period planets ever searched in a
GC.  The data are also of sufficient quality to be sensitive to
planets of smaller radii than any previous transit search in a GC.
From 3784 light curves extracted from the data, with a maximum observation
duration of \simm78 days, and using a BLS transit search followed by a
by-eye vetting, we identified 27 planet candidates in the data.  
An
analysis of the systematic noise in the light curves revealed that a
S/PN cutoff value of 12 should be used to remove probable false
positives, with only one of the planet candidates exceeding this
cutoff value, yet there still remains uncertainty as to whether this
might be a false alarm.  Despite this, information on this and six
other of our most promising candidates are presented.  The light curves
are publicly available at \cite{lightcurves}.

We calculated 3$\sigma$ occurrence rate upper limits based on a
non-detection of planets and occurrence rate ranges assuming our
S/PN$>12$ planet candidate as real, for a variety of period and planet
radius ranges.  Comparing these limits and rates to the literature,
for previous GC works, we find a factor of two lower occurrence rate
limit than was calculated by \cite{weldrake2008} for 
$\omega$ Cen for R$_p>1.5$ \rj\ objects.  We also improve on
the \cite{nascimbeni2012} limit for \simm1~\rj\ planets with
$\lesssim14$ day orbits, obtaining a 2$\sigma$ limit of $<0.93$\%. 
Our limit for a similar period
and radius range as the landmark study of \cite{gilliland2000} was
sensitive to, 1--8 days and 0.64--2 \rj, was ${<}2.1$\%, compared to the
${<}0.7$\% limit determined by \cite{masuda2017} 
using the \cite{gilliland2000} data.  Comparing with occurrence rates
calculated from field star transit surveys, our HJ occurrence rate
limits are factors of about four to six larger than the {\it Kepler} and
{\it TESS} rates.  Similarly, for RV studies, our HJ occurrence limits
are about a factor of two higher than the rate of \cite{wright2012}
and about a factor of three higher than the rate of \cite{mayor2011}.
Our rate upper limits for longer period orbits ($\gtrsim15$ days) of
\simm 1 \rj\ objects and for smaller planets (\simm 0.4 \rj\ and
larger) are much larger than the rates known for those regimes from
{\it Kepler} and are not very constraining, but are the first such limits
ever set for a GC.

Future work that could be done to build off these results includes photometric follow up of the planet candidates to
confirm the transits and improving the systematic noise characterization and
abatement  in the light curves to permit greater sensitivity
to lower S/PN transits.  Lowering the S/PN threshold would allow us to
put significantly better constraints on $P\lesssim8$ day planets with
$0.3~\mathrm{R}_\textrm{J}\lesssim \mathrm{R}_p \lesssim 0.6$ \rj\ planets and
for $0.6~\mathrm{R}_\textrm{J}\lesssim \mathrm{R}_p \lesssim 0.9$ \rj\ planets
across all periods examined.

\acknowledgements
We thank Kento Masuda for his
occurrence rate calculation, Waqas Bhatti for assistance with
\texttt{astrobase}, and Fei Dai and Josh Winn for providing a second vetting of our planet candidates and useful comments.  We also thank Wojtek Pych for assistance
in procuring and using the Cluster AgeS Experiment (CASE) light curves
for M4, though the data did not ultimately make it into the final work.
JH and GB acknowledge funding from NASA grant NNX17AB61G.
This research includes data collected by the {\it Kepler}/K2 mission and obtained from the MAST data archive at the Space Telescope Science Institute (STScI). Funding for the {\it Kepler} mission is provided by the NASA Science Mission Directorate. STScI is operated by the Association of Universities for Research in Astronomy, Inc., under NASA contract NAS 5–26555.
Support for MAST for non-HST data is
provided by the NASA Office of Space Science via grant NAG5-7584 and
by other grants and contracts. 
This research includes data from the European Space Agency (ESA)
mission {\it Gaia} (\url{https://www.cosmos.esa.int/gaia}), processed by
the {\it Gaia} Data Processing and Analysis Consortium (DPAC,
\url{https://www.cosmos.esa.int/web/gaia/dpac/consortium}). Funding
for the DPAC has been provided by national institutions, in particular
the institutions participating in the {\it Gaia} Multilateral
Agreement.
This research has made use of the SIMBAD database and the 
 VizieR catalogue access tool (DOI: 10.26093/cds/vizier), both
 operated at CDS, Strasbourg, France. The original description 
 of the VizieR service was published in \cite{vizier} and the
 description for SIMBAD was published in \cite{simbad}.
This research has made use of NASA's Astrophysics Data
System Bibliographic Services.

\facilities{{\it Gaia}, {\it Kepler}}

\software{\texttt{astrobase} \citep{astrobase}, 
  \texttt{astropy} \citep{astropy}, 
  \texttt{FITSH} \citep{pal}, 
  \texttt{k2mosaic} \citep{barentsen2016},
  \texttt{matplotlib} \citep{hunter2007},
  \texttt{numpy}  \citep{numpy},
  \texttt{scikit-learn} \citep{scikit-learn},
  \texttt{scipy} \citep{scipy}, 
  \texttt{VARTOOLS} \citep{vartools}
}

\bibliographystyle{aasjournal}

\end{document}